\documentclass[fleqn,twoside]{article}
\usepackage{espcrc2}

\usepackage{graphicx}

\usepackage{epsfig}
\usepackage[figuresright]{rotating}

\newcommand{\AmS}{{\protect\the\textfont2
  A\kern-.1667em\lower.5ex\hbox{M}\kern-.125emS}}

\title{Neutrino Factory Designs and R\&D}

\author{Steve Geer\address[MCSD]{Fermi National Accelerator Laboratory, 
P.O. Box 500, Batavia, IL 60510, USA}%
        \thanks{
Talk presented at the XXth International Conference on Neutrino Physics 
and Astrophysics, May 25-30, 2002, Munich, Germany. 
This talk is based upon the work of many people. My own 
humble contributions were 
supported at the Fermi National Accelerator Laboratory, which is operated 
by Universities Research Association, under contract No. DE-AC02-76CH03000 
with the U.S. Department of Energy.}
        }
       
\begin{document}

\begin{abstract}
European, Japanese, and US Neutrino Factory designs are presented. 
The main R\&D issues, and the associated R\&D programs, are discussed.
\vspace{1pc}
\end{abstract}

\maketitle

\section{Introduction}
\label{Intro}

The development of a 
very intense muon source capable of producing a millimole of muons per 
year would enable a Neutrino Factory~\cite{sgprd}, 
and perhaps eventually a Muon Collider~\cite{mucollider}, to be built. 
In the last two years Neutrino Factory physics studies~\cite{nufact-physics} 
have mapped out an exciting Neutrino Factory physics program. 
In addition, Neutrino Factory feasibility studies~\cite{study1,study2,studyJ}  
have yielded designs that appear to be ``realistic'' provided the 
performance parameters for the critical components can be achieved. 
Some of the key components will need a vigorous R\&D program to meet the 
requirements. Neutrino Factory R\&D activities in Europe~\cite{europe-RD}, 
Japan~\cite{studyJ,japan-RD}, and the US~\cite{us-RD} have resulted in 
three promising variants of the basic Neutrino Factory design. 
In the following the various Neutrino Factory schemes are briefly described. 
The main R\&D issues and the ongoing R\&D programs are summarized.

\section{Neutrino Factory Schemes}

In all of the present Neutrino Factory schemes  
an intense multi-GeV proton source is used to make 
low energy charged pions which are confined within  
a large acceptance decay channel. The daughter muons 
produced from $\pi^\pm$ decays are also confined within the channel. 
However, the muons occupy a large phase-space volume which 
presents the main challenge in designing a Neutrino Factory. 
In the US and European designs the strategy is to first 
reduce the energy spread of the muons by  manipulating the 
longitudinal phase-space they occupy using a technique 
called ``phase rotation''. 
The transverse phase-space occupied by the muons is then reduced 
using ``ionization cooling~\cite{cool}''. After phase rotation and 
ionization cooling the resulting muon phase space fits within 
the acceptance of a normal type of accelerator. In the Japanese scheme 
an alternative strategy is pursued in which the large muon phase-space 
is accommodated using so called FFAG's, which are very large acceptance 
accelerators. Finally, in all three schemes the muons are accelerated 
to the desired final energy (typically in the range from 20 - 50~GeV), 
and injected into a storage ring with either two or three long straight 
sections. Muons decaying within the straight sections produce intense 
neutrino beams. If the straight section points downwards, the resulting 
beam is sufficiently intense to produce thousands of neutrino interactions 
per year in a reasonably sized detector on the other side of the 
Earth~\cite{sgprd}~!

\subsection{US Scheme}

In the last 2 years there have been two Neutrino Factory ``Feasibility'' 
Studies~\cite{study1,study2} in the US. Within these studies  
engineering designs have been developed and detailed simulations 
performed for each piece of the Neutrino Factory complex. 
Study I was initiated by the Fermilab Director, and 
conducted between October 1999 and April 2000. 
\begin{figure}
\centering
\epsfxsize260pt
\epsffile{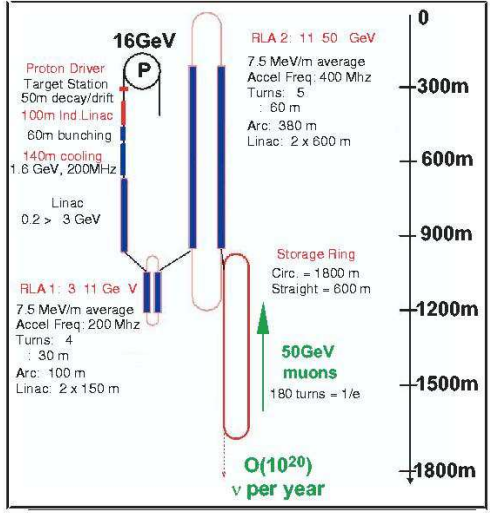}
\vspace{-1.0cm}
\caption{US Neutrino Factory schematic from design study I~[4].}
\label{fig:US}
\end{figure}
The Study I design (Fig.~\ref{fig:US})  
was for a 50~GeV Neutrino Factory with the neutrino beams pointing 
13$^\circ$ below the horizon ($L = 2900$~km). 
The proton source~\cite{pstudy} consisted of 
a 16~GeV synchrotron producing 3~ns long bunches at 15~Hz, 
and providing a 1.2~MW beam ($4.5 \times 10^{14}$ protons per sec) on 
an 80~cm long carbon target located within a 20~T solenoid. The solenoid 
radially confines nearly all of the produced $\pi^\pm$. Downstream 
of the target the pions propagate down a 50~m long 
decay channel consisting of a 1.25~T super-conducting solenoid which 
confines the $\pi^\pm$ and their daughter muons within a warm bore of 60~cm. 
At the end of the channel 95\% of the initial $\pi^\pm$ have decayed
and, per incident proton, there are $\sim 0.2$ muons with energies $< 500$~MeV 
captured within the beam transport system.  
The muon system downstream of the decay channel is designed to produce 
cold muon bunches with a central momentum of 200~MeV/c. 
Beyond the decay channel, the muon energy 
spread is reduced using a 100~m long induction linac 
to accelerate the late low 
energy particles and deaccelerate the early high energy particles 
(phase rotation). 
A 2.45~m long liquid hydrogen absorber is then 
used to lower the central energy of the muons. 
Muons with energies close to the central value are then captured 
within bunches using a 201~MHz RF system within a 60~m long channel. 
Throughout the induction linac, liquid hydrogen absorber, 
and bunching system the muons are confined 
radially using a solenoid field of a few Tesla. 
The buncher produces a string of bunches, captured 
longitudinally, but still occupying a very large transverse phase-space.  
Downstream of the buncher the muons are cooled transversely within a 
120~m long ionization cooling channel. The phase-space occupied by 
the muon bunches will then fit with the acceptance of an acceleration 
system consisting of a linac that accelerates the muons to 3~GeV, and two 
recirculating linear accelerators (RLA's) that raise the muon energies 
to 50~GeV. The muons are then injected into a storage ring which has a 
circumference of 1800~m, and has two 600~m long straight sections. 
One third of the injected muons will decay in the downward pointing 
straight section. A detailed simulation of the Study I design shows that
this scheme will produce about $6 \times 10^{19}$ muon decays per 
operational year (defined as $2 \times 10^7$ seconds) 
in the downward pointing straight-section, which is a factor of 3 less 
than the initial design goal for the study. The resulting beam intensity 
would be sufficient for a so-called ``entry-level'' Neutrino Factory, 
but insufficient for a ``high-performance'' machine. 

Study II, initiated by the BNL Directorate, built upon the work done 
during Study I and focussed 
on improving the Neutrino Factory design to achieve higher beam intensities. 
The main improvements came from using 
(i) a 4~MW proton source, 
(ii) a liquid Hg target, 
(iii) an improved induction linac design, 
and (iv) an improved cooling channel design. 
Detailed simulations of the Study II design predicted $2 \times 10^{20}$ 
muon decays per operational year (redefined as $1 \times 10^7$ seconds) 
in the downward pointing straight-section, thus achieving the initial 
goal. 
\begin{figure}
\epsfxsize190pt
\epsffile{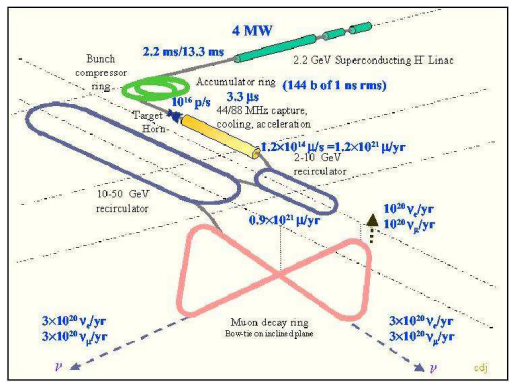}
\vspace{-0.5cm}
\caption{European Neutrino Factory scheme~[12].}
\label{fig:CERN}
\end{figure}

\subsection{European Scheme}

European and US Neutrino Factory designs have many similarities. 
However, European studies~\cite{cern} (Fig.~\ref{fig:CERN}) 
have explored alternative 
technologies for several subsystems: 
(i) The proton driver consists of a 4~MW 2.2~GeV SC linac  
with an accumulator ring to produce short pulses~\cite{europe-p}. 
(ii) In addition to a liquid Hg jet, a water 
cooled Ta sphere target is being considered. 
(iii) The charged pions are focussed using a magnetic horn (rather than 
a high-field solenoid) with a 4~cm waist radius, and a peak current 
of 300~kA. 
(iv) After a 30~m drift, the muon energy spread is reduced using 44~MHz 
RF cavities (rather than an induction linac).
(v) The ionization cooling channel uses 44~MHz and 88~MHz RF
cavities, a lower frequency cavities than employed in the US scheme. 
(vi) A bowtie-shaped storage ring (rather than a race-track design) 
has been considered for the final muon ring. 
Simulations of the European design predict that the resulting neutrino 
beams will have intensities comparable to the corresponding beams from 
the US design. A comprehensive design study is not yet complete.
\begin{figure}
\centering
\epsfxsize190pt
\fbox{
\epsffile{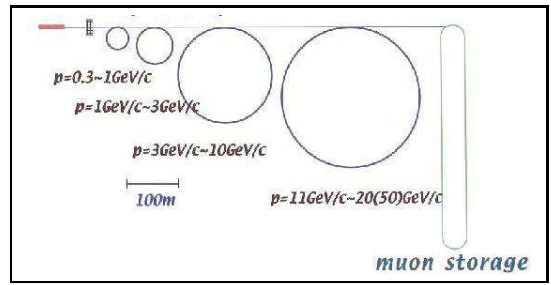}
}
\vspace{-0.5cm}
\caption{Japanese Neutrino Factory schematic~[8].}
\label{fig:Japan}
\end{figure}

\subsection{Japanese Scheme}

The front end of the Japanese design (Fig.~\ref{fig:Japan}) 
consists of a 4~MW 50 GeV proton synchrotron  
(corresponding to an upgraded JHF complex) producing a beam incident on a 
target within a 12~T solenoid, followed by a 
large acceptance pion decay channel. Instead of manipulating and 
cooling the phase space occupied by the muons exiting the decay channel, 
very large acceptance FFAG (Fixed Field Alternating Gradient) accelerators 
are used to raise the beam energy before injecting into a storage ring 
with long straight sections. The predicted neutrino yield from this scheme 
seems to be comparable to the corresponding predicted yields from the 
US/European designs. 
Although the Japanese Neutrino Factory scheme might benefit from the addition 
of some muon cooling, in principle the use of FFAGs evades the 
need to cool the muons before injecting them into an accelerator. This 
simplification comes at the price of a more challenging accelerator, 
requiring complicated large aperture magnets and broad-band low frequency 
high gradient RF cavities. Modern design 
tools have made practical the task of designing the FFAG magnets, and 
a small proof-of-principle (POP) FFAG accelerator~\cite{pop} has been 
built and successfully operated. 
A second test FFAG, designed to accelerate protons to 150~MeV, 
is under construction.
Furthermore, with some US participation, 
an R\&D program is underway in Japan to develop the required cavities. 
It is too early to conclude whether the promising Japanese Neutrino Factory 
scheme will lead to a more cost effective solution than the European/US-type 
designs. It may even be that the optimum solution consists of 
a combination of the two concepts, with some 
phase space manipulation and cooling, but using large acceptance FFAGs 
for the acceleration.

\section{Pion production \& Target R\&D}

\begin{figure}
\centering
\epsfxsize200pt
\epsffile{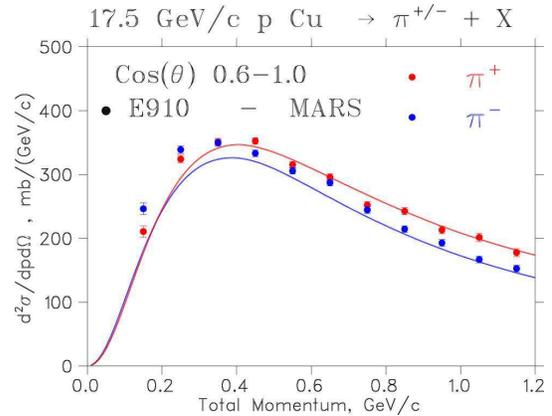}
\vspace{-0.5cm}
\caption{Pion production measurements from the BNL E910 experiment, compared 
with MARS predictions~[15].}
\label{fig:E910}
\end{figure}

To produce a sufficient number of $\pi^\pm$, all Neutrino Factory 
schemes begin with a MW-scale proton driver.
\begin{figure}
\centering
\epsfxsize200pt
\epsffile{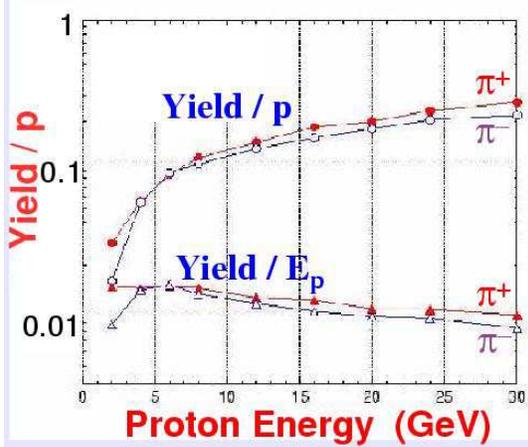}
\vspace{-0.5cm}
\caption{MARS predictions~[4]. 
The number of $\pi^+ + \mu^+$ (filled symbols) 
and $\pi^- + \mu^-$ (open symbols) within an energy interval 
30~MeV $< E < 230$~MeV is shown 9~m downstream of an 
80~cm long 0.75~cm radius carbon 
target within a 20~T solenoid, and tilted 50~mrad with respect to the solenoid 
axis. Triangles show the yields divided by the primary proton energy.
}
\label{fig:yields}
\end{figure}
Figure~\ref{fig:yields} shows predicted yields for $\pi^\pm$ captured 
within a decay channel downstream of a Neutrino Factory target system. 
Over a broad interval of beam energies, at fixed beam power the 
yields are approximately independent of beam energy. 
The E910 experiment at BNL~\cite{E910} has recently measured $\pi^\pm$  
yields for several targets and different incident proton beam energies. Some  
of the results are shown in Fig.~\ref{fig:E910}. The measurements 
are in fair agreement with MARS calculations~\cite{E910}. The pion yields 
peak in the region 300 - 500~MeV/c. Hence, Neutrino Factory designs 
tend to have $\pi^\pm$ collection systems optimized to capture particles 
with momenta in this range. 
In the next few years we can anticipate further particle production 
measurements from the HARP 
experiment~\cite{harp} at CERN, and the Fermilab E907 experiment~\cite{P907}.

Entry-Level Neutrino Factories providing a few $\times 10^{19}$ useful 
muon decays per year require proton beams with beam powers of $\sim 1$~MW, 
and short proton bunches, typically a few ns long. It is believed that 
carbon targets can operate with these beam parameters. This has been tested 
at BNL by the E951 Collaboration~\cite{target}. 
Two different types of carbon rod were 
exposed to the AGS beam and strain gauge data taken. The beam induced 
longitudinal pressure waves and transverse reflections were both measured. 
The strains in the two types of rod differed by an order of magnitude, 
the most promising rod being made from an anisotropic carbon-carbon 
composite.

High performance Neutrino Factories providing O($10^{20}$) useful 
muon decays per year require $\sim 4$~MW proton beams.
Solid targets will melt 
unless very efficient cooling strategies can be developed. Rotating metal 
bands~\cite{band_target} and water cooled Ta spheres~\cite{sphere_target} 
are being considered, but the presently favored 
solution is to avoid problems with target melting and integrity by using 
a liquid Hg jet. In the US and Japanese schemes the target is in a high-field 
solenoid. Hence, the main R\&D issues are (i) can a Hg jet be injected within 
a high-field solenoid without magneto-hydrodynamic effects disrupting the 
jet, (ii) after the jet has been destroyed by one proton pulse will it 
re-establish itself before the next pulse, and (iii) can the disrupted 
jet be safely contained within the target system.
\begin{figure}
\centering
\epsfxsize100pt
\epsffile{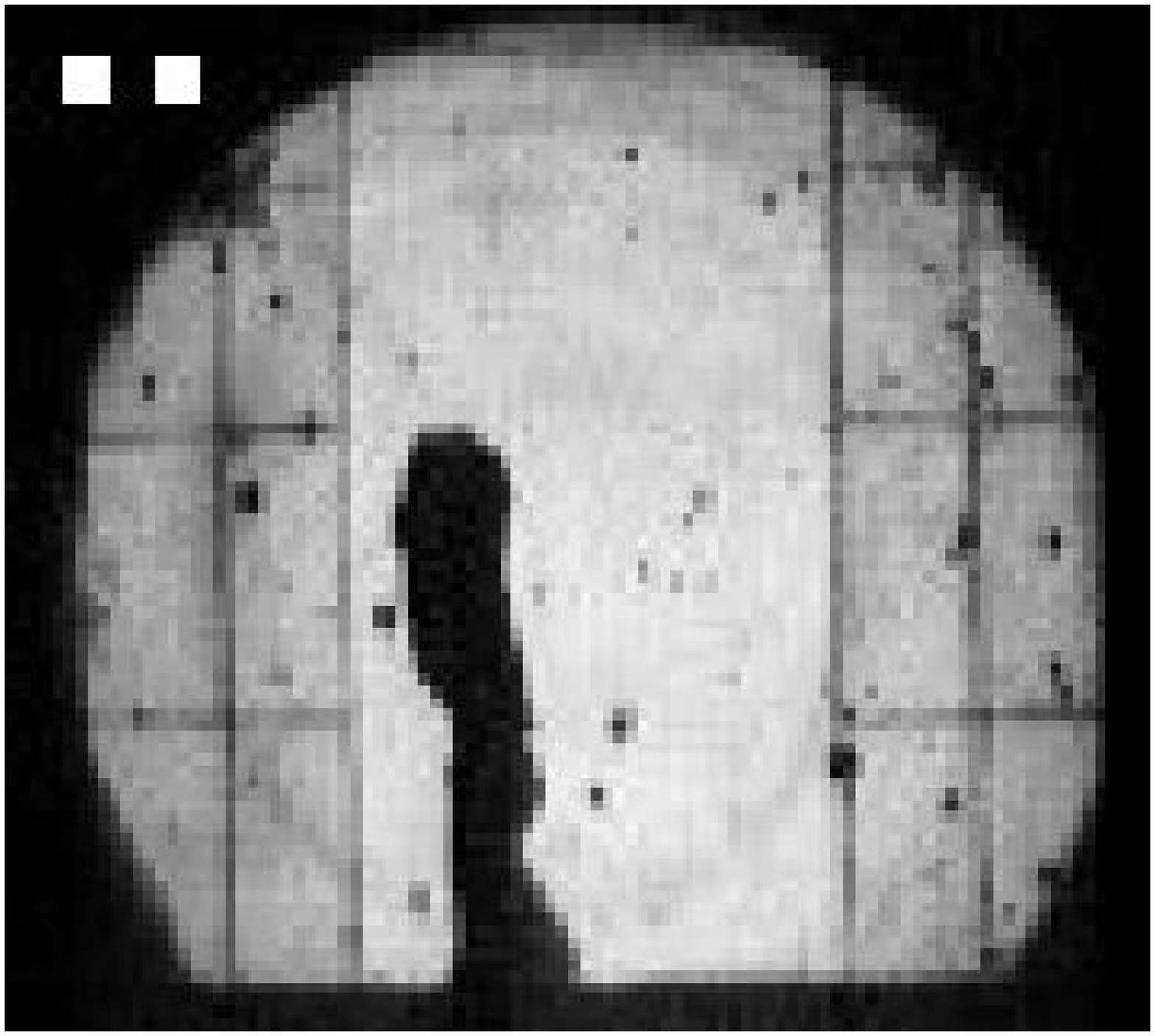}
\epsfxsize97pt
\epsffile{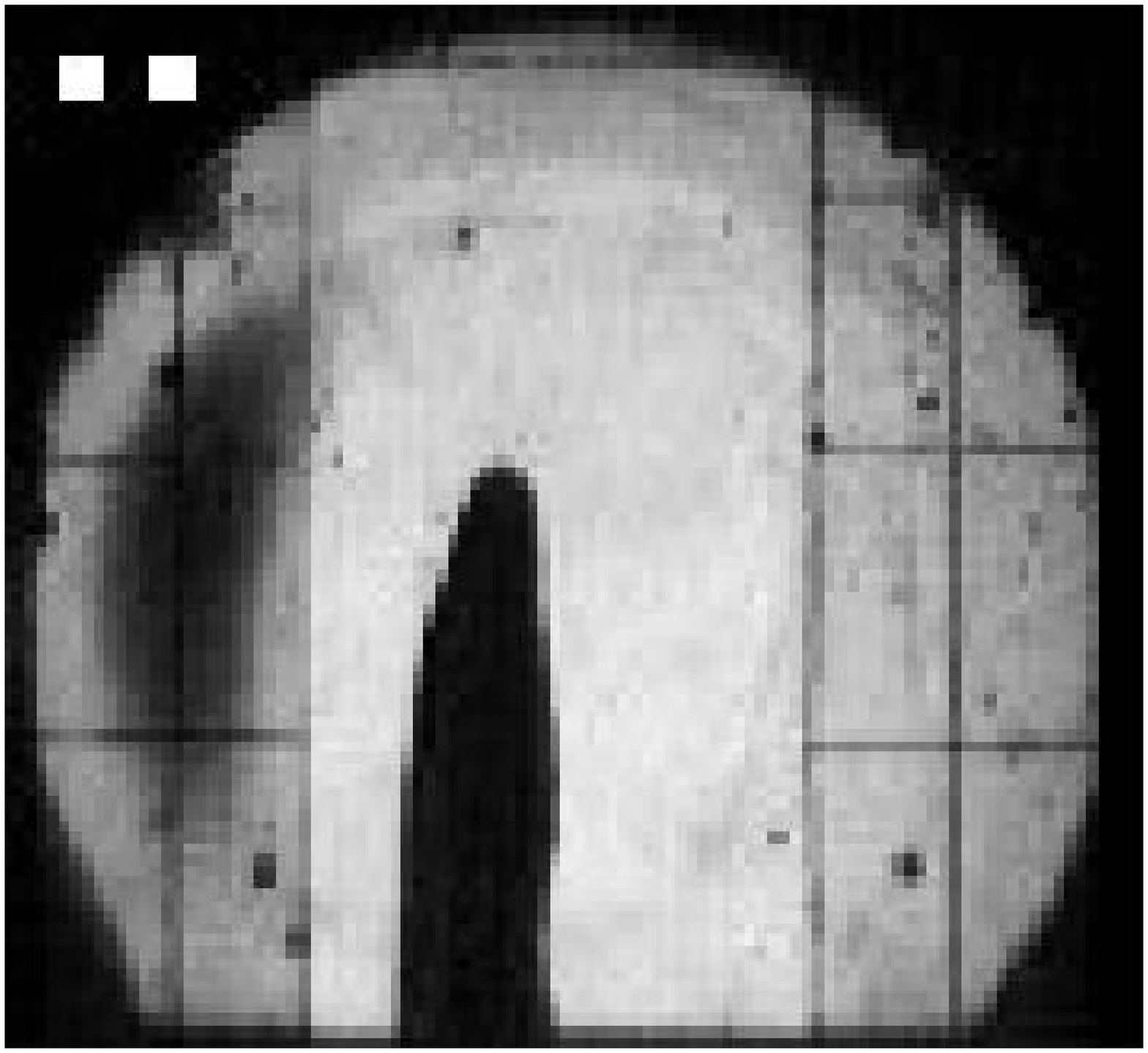}
\vspace{-0.5cm}
\caption{CERN-Grenoble tests in which a liquid Hg jet is injected 
into a solenoid providing 0 field (top picture) and 13~T 
(bottom picture)~[13].}
\label{fig:grenoble}
\end{figure}
\begin{figure}
\centering
\epsfxsize220pt
\epsffile{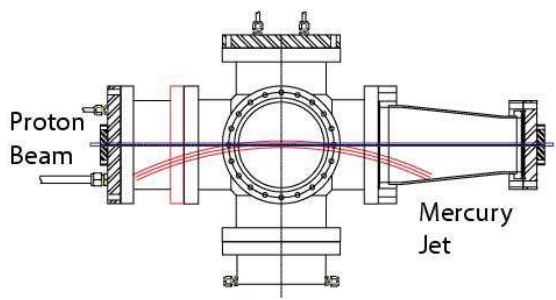}
\vspace{-0.8cm}
\caption{Schematic of the E951 system to test a Hg jet in the BNL AGS 
beam~[19].}
\label{fig:E951}
\end{figure}
\begin{figure*}
\centering
\epsfxsize30pc
\epsffile{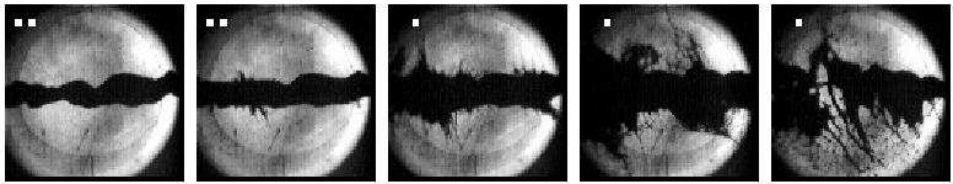}
\vspace{-0.5cm}
\caption{BNL E951 results~[19]. 
The sequence of pictures shows the Hg jet at the 
time of impact (t=0) of the AGS beam, and at t = 0.75~ms, 2~ms, 7~ms, 
and 18~ms. Time increases from left to right.}
\label{fig:e951-results}
\end{figure*}
To address these questions, R\&D has begun at CERN and 
BNL. Photographs of a liquid Hg jet injected into 
a 13~T solenoid at Grenoble are shown in Fig.~\ref{fig:grenoble}. 
The solenoid field seems to damp surface tension waves, 
improving the jet characteristics. 
The BNL E951 liquid Hg jet setup~\cite{target} in the 24~GeV AGS beam is 
shown in 
Fig.~\ref{fig:E951}. Figure~\ref{fig:e951-results} shows a sequence of 
pictures of the 1~cm diameter Hg jet taken by a high speed camera from the 
time of impact of $4 \times 10^{12}$ protons in a 150~ns long pulse 
from the AGS. Jet dispersal is delayed for about $40\mu$s. 
It takes several ms before the jet, which has an 
initial velocity of 2.5~m/sec, is disrupted. The velocity of 
the out-flying Hg filaments appears to scale with beam intensity, 
and is $\sim 10$~m/sec for a deposited energy of 25~J/g. 
These velocities are modest, implying that the disrupted jet can 
be easily contained within the target system. Furthermore, the Hg jet 
dispersal is mostly in the transverse direction, and after disruption 
it has been found that the jet quickly re-establishes itself. 
However, high performance 
Neutrino Factories will require Hg jets with higher velocities 
($\sim 20$~m/sec). The next steps in the E951 R\&D  
require beam tests with a factor of a few higher beam intensities, and 
finally beam tests in which the Hg jet is injected into a 20~T solenoid.

\section{Muon Cooling R\&D}

Before the muons can be accelerated   
the transverse phase-space they occupy must be reduced 
so that the muon beam fits within the acceptance of an accelerator. 
This means the muons must be ``cooled'' by at least a factor of a few in each 
transverse plane, and this must be done fast, before the muons decay. 
Stochastic-- and electron--cooling are too slow. 
It is proposed to use a new cooling technique, namely 
``ionization cooling''~\cite{cool}. 
In an ionization cooling channel the muons 
pass through an absorber in which they lose transverse-- and 
longitudinal--momentum by {\it dE/dx} losses. The longitudinal momentum 
is then replaced using an RF cavity, and the process is 
repeated many times, reducing the transverse momenta. This  
cooling process will compete with transverse heating due to 
Coulomb scattering. To minimize the effects of scattering we chose 
low--Z absorbers placed in the cooling channel 
lattice at positions of low--$\beta_\perp$ so that the typical radial 
focusing angle is large. If the focusing angle is much larger 
than the average scattering angle then scattering will not have much 
impact on the cooling process. 
In the US, detailed simulation studies have identified two promising 
linear cooling channel designs: ``SFOFO'' and ``DFLIP''. 
\begin{figure*}
\centering
\epsfxsize25pc
\epsffile{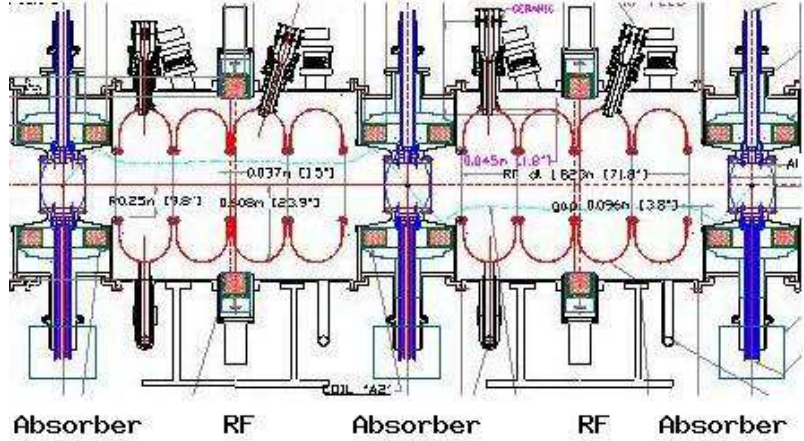}
\vspace{-0.7cm}
\caption{SFOFO cooling channel design~[5]. A 5.5~m long section 
is shown, consisting of two 
200 MHz four-cell cavities interleaved with three liquid 
hydrogen absorbers.}
\label{fig:sfofo}
\end{figure*}

In the SFOFO lattice the absorbers are 
placed at low--$\beta_\perp$ locations 
within high-field solenoids. The field rapidly decreases from a 
maximum to zero at the absorber center, and then increases to a maximum 
again with the axial field direction reversed. 
Figure~\ref{fig:sfofo} shows the design for a 5.5~m long section of the 
$\sim 100$~m long cooling channel. The section shown has 30~cm long 
absorbers with a radius of 15~cm, within a system of solenoids with 
a peak axial field of 3.5~T. 
Towards the end of the cooling channel the maximum field is 
higher (5~T) and the lattice period shorter (3.3~m).  
The RF cavities operate at 201~MHz and provide a peak gradient of 17~MV/m. 
Detailed simulations predict that the SFOFO channel increases  
the number of muons within the accelerator acceptance  
by a factor of 3-5 (depending 
on whether a large- or very-large acceptance accelerator is used).

In the DFLIP lattice the solenoid field remains constant
over large sections of the channel, reversing direction only twice. 
In the early part of the channel the muons lose mechanical angular momentum 
until they are propagating parallel to the axis. After the first field 
flip the muons have, once again, mechanical angular momentum, and 
hence move along helical trajectories with Lamour centers along the 
solenoid axis. Further cooling removes the mechanical angular momentum, 
shrinking the beam size in the transverse directions. 
The field in the early part of the channel is 3~T, increasing to 7~T for
the last part. Detailed simulations 
show the performances of the DFLIP and SFOFO channels are comparable.

Recently there has been considerable progress in the design of 
cooling channels configured in a ring, and incorporating wedge-shaped 
absorbers at locations where higher energy muons pass through the thick end 
of the wedge. This cools both longitudinal and transverse 
phase space. 
The simulations, which predict ring coolers have wonderfull 
performance, do not yet incorporate realistic absorbers or detailed 
injection and extraction schemes. It remains to be seen if realistic 
ring coolers will result in greatly improved performance and significant 
cost savings compared to the present baseline cooling channels. 

\subsection{MUCOOL R\&D}

Muon cooling channel design and development is being pursued within the 
US by the MUCOOL collaboration~\cite{MUCOOL}. The MUCOOL mission 
includes bench--testing all cooling channel components and  
beam--testing a cooling channel section. The main component issues 
that must be addressed are 
(i) can sufficiently high gradient RF cavities be built and operated 
in the appropriate magnetic field and radiation environment, and 
(ii) can liquid hydrogen absorbers with thin enough windows be built 
so that the $dE/dx$ heating can be safely removed. The MUCOOL collaboration 
has embarked on a design--, prototyping--, and testing--program that  
addresses these questions.

\subsection*{RF Cavity Development}

\begin{figure}
\centering
\epsfxsize150pt
\epsffile{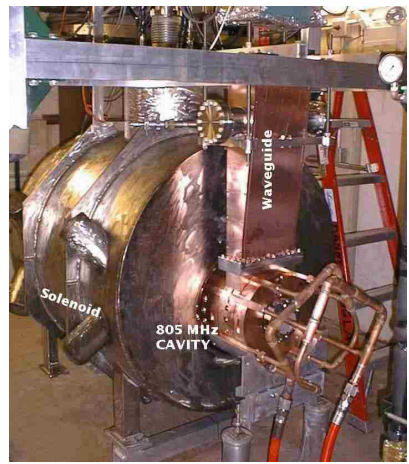}
\vspace{-0.7cm}
\caption{High power 805 MHz cavity test in the MUCOOL Lab G test 
area, Fermilab~[25].}
\label{fig:labg}
\end{figure}

Cooling channel studies initially began with Muon Collider 
cooling channel designs using 805~MHz cavities operating in a 5T 
solenoid, and providing a peak gradient on axis   
of $\sim30$~MV/m. This deep potential well is needed to keep the 
muons bunched as they propagate down the channel. 
This requirement led to two cavity concepts: 
(a) an open cell design, and (b) a design in which the 
penetrating nature of the muons is exploited by closing the RF aperture 
with a thin conducting Be window (at fixed peak power this doubles 
the gradient on axis). 
The MUCOOL collaboration has pursued an aggressive 805 MHz cavity 
development program. 
A 12~MW high power test facility has been built and operated at 
Fermilab (Lab G), enabling 
805 MHz cavities to be tested within a 5T solenoid. 
The main results to date are:  
(i) An open cell cavity suitable for a muon cooling channel has been 
designed, an aluminum model built and measured, and a prototype copper 
cavity built, tuned, and tested at full power (Fig.~\ref{fig:labg}). 
Dark current produced by the cavity has 
been identified as an important R\&D issue~\cite{dark_current}. 
(ii) A Be foil cavity has been designed at LBNL, a low power test cavity 
built and measured, and foil deflection studies made~\cite{BeWindows} 
to ensure the cavity 
does not detune when the foil is subject to RF heating. 
The foil deflection is reasonably well understood for small displacements. 
A high power copper 
cavity with thick Be windows has been built at LBNL and the 
University of Mississippi, and tested at full power at Lab G. Magnetic field 
studies will start soon. 

The cooling channel designs developed in the US Neutrino Factory studies 
require 201~MHz RF cavities providing $\sim 17$~MV/m on axis. 
Preliminary cavity designs have been made. There are two concepts, 
both of which close the cavity aperture. The options are to use (a) a thin 
Be foil, exploiting the work done for the 805 MHz cavity, or 
(b) use a grid of hollow conducting tubes. Preliminary mechanical tests 
for both the grid and foil concepts are planned. 
The 805~MHz R\&D program will be 
extended so that windows and grids can be tested at high power. The 
805~MHz program will also enable us to obtain 
a better understanding of how to build low 
dark-current NCRF cavities, providing critical guidance for 
the 201~MHz cavity prototype that we hope to build and test in the next 
couple of years.
\begin{figure}
\centering
\epsfxsize9pc
\epsffile{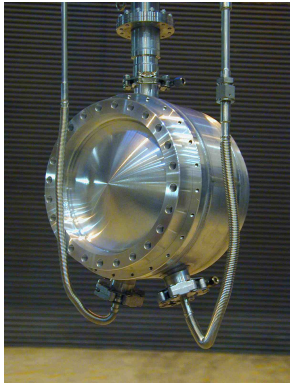}
\vspace{-0.5cm}
\caption{KEK prototype absorber. The liquid hydrogen is to be mixed 
by convection and cooled with a local heat exchanger.}
\label{fig:convection}
\end{figure}

\subsection*{Absorber Development}

The cooling channel liquid hydrogen absorbers must have 
very thin windows to minimize multiple scattering, and must tolerate 
heating of O(100~W) from the ionization energy deposited by the 
traversing muons. Typical absorber parameters are: 
35~cm long, 18~cm radius, with 360~$\mu$m thick Al windows, operated 
at 1.2~atm. 
To adequately remove the heat from the LH$_2$ requires good 
transverse mixing. There are two absorber designs that are 
being pursued~\cite{Dan}: (i) Forced flow design. The LH$_2$ is 
injected into the 
absorber volume through nozzles, and cooled using an external 
loop and heat exchanger. (ii) Convection design. Convection 
is driven by a heater
at the bottom of the absorber volume, and heat removed by a heat exchanger
on the outer surface of the absorber. A forced flow absorber 
prototype is being designed at the Illinois Institute of Technology (IIT) 
and will be constructed in the coming year. A convection prototype, 
designed by IIT, KEK, and the University of Osaka, is being constructed 
in Japan (Fig.~\ref{fig:convection}). 
Both absorbers will be tested at Fermilab.

Prototype 15~cm radius aluminum absorber windows have been made at the 
University of Mississippi, and pressure tested at Northern Illinois 
University~\cite{Dan}. The window thickness 
and profile were measured at FNAL and found 
to be within 5\% of the nominal envelope. Strain gauge and photogrammetric 
measurements were made as a function of pressure, and the results compared 
with FEA predictions. The last window was tested at LN$_2$ temperature. The 
FEA predictions give an excellent description of the deformation with 
pressure and the rupture pressure. 
\begin{figure}
\centering
\epsfxsize17pc
\epsffile{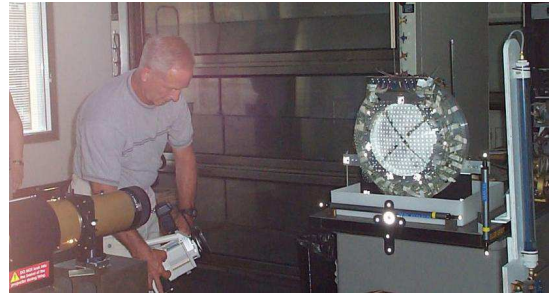}
\vspace{-0.5cm}
\caption{Absorber window test, showing an array of dots 
projected onto the window for photogrammetric measurements of its 
shape as it deforms under pressure~[27].}
\label{fig:absorber}
\end{figure}

\subsection{European Cooling R\&D Program}

The European cooling channel design~\cite{Alessandra} 
is similar to the US design, 
but is based on 44~MHz and 88~MHz cavities~\cite{cern_cavities} 
rather than 201~MHz cavities. 
To minimize the radii of the solenoids used to confine the muons 
the cavities have been designed to wrap around 
the solenoids. 
The initial transverse cooling is performed using 44~MHz cavities with 
four 1~m long RF cells between each 24~cm long LH$_2$ absorber. 
The beam is then accelerated from 200~MeV to 300~MeV, and the cooling 
is continued using 88~MHz cavities with eight 0.5~m long cells 
between each 40~cm long LH$_2$ absorber. 
Simulations of the channel performance 
with detailed field-maps have not yet been made. However, simulations 
using simpler field maps yield promising results: the effect of the 
channel is to increase the number of muons within the acceptance of the 
subsequent accelerating system by a factor of about 20. Whether or not 
the predicted  
increased yield is significantly degraded when full simulations are 
performed remains to be seen. In the meantime, a prototype 88~MHz cavity 
is being prepared at CERN 
for high power tests within the coming year. 

\subsection{Cooling Experiments}

A sequence of muon cooling-related experiments is being planned. 
The first, the MUSCAT experiment, 
is already under way at TRIUMF. 
The second, the MUCOOL Component Test Experiment, is under 
construction at the Fermilab Linac. The third, an International 
Cooling Experiment~\cite{MICE}, is in the planning stage.

The goal of the MUSCAT experiment is the precise measurement of 
low energy (130, 150, and 180~MeV/c) 
muon scattering in a variety of materials that might be  
used in a cooling channel. In a second phase MUSCAT will  
measure straggling. Scattering measurements for Li, Be, 
C, Al, CH$_2$, and Fe have already been made.

A MUCOOL test area located at the end of the Fermilab 400~MeV Linac 
is under construction. 
The project is being pursued in two phases. In Phase 1 a LH$_2$ absorber 
test facility is being built, which will enable the first prototype 
absorbers to be filled. In Phase 2  
a linac beam will be brought to the absorber area, and the 5T solenoid 
will be moved from Lab G so that the absorber can be tested in a magnet 
whilst exposed to a proton beam. The beam intensity and spot size 
will be designed to mimic the total ionization energy deposition 
and profile corresponding to the passage of $10^{12} - 10^{13}$ muons 
propagating within a cooling channel. In addition, 201~MHz RF power 
will be piped to the test area from a nearby test-stand, enabling 
high-power tests to be made of a prototype 201~MHz cooling channel 
cavity.

A Europe-Japan-US International Cooling Experiment (MICE) is currently being 
planned~\cite{MICE}. The goals are to (i) place a cooling channel section in a muon 
beam, and (ii) demonstrate our ability to precisely simulate the passage 
of muons confined within a periodic lattice as they pass through 
LH$_2$ absorbers and high-gradient RF cavities. In the envisioned 
experiment muons are measured one at a time at the input and output of the 
cooling section, and the precise response of the muons to the cooling 
section is determined. The main challenge to the design of this type 
of experiment arises from the prolific X-ray and dark current 
environment created by the 
RF cavities. This is currently under study at Lab~G and elsewhere. 
The MICE Collaboration has been invited to submit a full proposal 
to the Rutherford Lab at the end of 2002.

\section{Muon Acceleration}

The acceleration system has been identified as one of the cost drivers 
for a Neutrino Factory. In the US and European schemes the main acceleration 
systems use SC cavities. The US scheme uses 201~MHz SCRF 
delivering gradients of 15~MV/m with Q~$\sim 5 \times 10^9$. Note that 
201~MHz is a relatively 
low frequency and the cavities are therefore large. 
The associated R\&D issues are related 
to microphonics, fabrication and cleaning techniques and, because of the 
large stored energy, quench protection. Furthermore, the cavities must 
tolerate whatever stray magnetic fields they see within the accelerating 
lattice. To address these issues a 201~MHz SC cavity has been constructed 
at CERN and sent to Cornell for high-power testing.

\section{Summary and Prospects}

Muon sources capable of delivering O($10^{20}$) muons per year 
seem feasible, and would enable Neutrino Factories, and perhaps 
eventually Muon Colliders, to be built. Neutrino Factory designs 
have been developed in Europe, Japan, and the US, and a healthy R\&D program 
is underway. The most challenging R\&D 
questions are associated with targets for MW-scale proton beams, 
and the development of an ionization cooling channel.
With the present level of support, we can expect much progress 
in Neutrino Factory R\&D over the next few years. However, 
the news is not all good. The future 
level of support is uncertain. I believe 
a significant increase (factor of two ?) will 
be needed, sustained over a handful of years, if we are ever to 
arrive at a ``Technical Design Report''. In addition, there must be 
good international collaboration to enable the most promising design 
to be eventually chosen. There is already a healthy 
dialogue between the European, Japanese, and US R\&D teams, 
and some cross-participation in the various R\&D programs. 
Finally, it should be noted that Neutrino Factory R\&D is being pursued 
by engineers, accelerator physicists, and particle physicists 
from Laboratories and Universities in Europe, Japan and the US. There 
are a broad range of interesting sub-projects to be pursued. With 
adequate support, the prospects seem bright.

\end{document}